\documentclass[11pt]{article}
\usepackage{lmodern}
\usepackage{dsfont}
\usepackage{environ}
\input{definitions.sty}

\newcommand{\pmin}{{p_\text{min}}}
\newcommand{\pmax}{{p_\text{max}}}

\newcommand{\xs}{\ensuremath{x^\star}}
\newcommand{\ys}{\ensuremath{y^\star}}
\newcommand{\cm}{\ensuremath{\mathsf{CM}}\xspace}
\newcommand{\fee}{\ensuremath{\mathsf{fee}}\xspace}
\newcommand{\LVR}{\ensuremath{\mathsf{LVR}}\xspace}

\newcommand{\N}{{\mathbb{N}}}

\newcommand{\calI}{\mathcal{I}}

\newcommand{\calL}{\mathcal{L}}

\def\<{\langle}
\def\>{\rangle}

\newcommand{\emailhref}[1]{\href{mailto:#1}{\tt #1}}

\begin{document}

\title{FLAIR: A Metric for Liquidity Provider Competitiveness \\ in Automated Market Makers\footnote{Working paper.}}

\author{
 	    \textbf{Jason Milionis} \\
        \small Department of Computer Science \\
 		\small Columbia University \\
        \small Research, Uniswap Labs \\
 		\small \emailhref{jm@cs.columbia.edu}
        \and
        \textbf{Xin Wan} \\
        \small Research \\
 		\small Uniswap Labs \\
 		\small \emailhref{xin@uniswap.org}
        \and
        \textbf{Austin Adams} \\
        \small Research \\
 		\small Uniswap Labs \\
 		\small \emailhref{austin@uniswap.org}
}
\date{}
\maketitle

\thispagestyle{empty}

\begin{abstract}
This paper aims to enhance the understanding of liquidity provider (LP) returns in automated market makers (AMMs).
LPs face market risk as well as adverse selection due to risky asset holdings in the pool that they provide liquidity to and the informational asymmetry between informed traders (arbitrageurs) and AMMs.
Loss-versus-rebalancing (LVR) quantifies the adverse selection cost \parencite{jason_lvr}, and is a popular metric to evaluate the flow toxicity to an AMM.
However, individual LP returns are critically affected by another factor orthogonal to the above: the competitiveness among LPs.
This work introduces a novel metric for LP competitiveness, called FLAIR (short for fee liquidity-adjusted instantaneous returns), that aims to supplement LVR in assessments of LP performance to capture the dynamic behavior of LPs in a pool.
Our metric reflects the characteristics of fee return-on-capital, and differentiates active liquidity provisioning strategies in AMMs.
To illustrate how both flow toxicity, accounting for the sophistication of the counterparty of LPs, as well as LP competitiveness, accounting for the sophistication of the competition among LPs, affect individual LP returns, we propose a quadrant interpretation where all of these characteristics may be readily visualized.
We examine LP competitiveness in an ex-post fashion, and show example cases in all of which our metric confirms the expected nuances and intuition of competitiveness among LPs.
FLAIR has particular merit in empirical analyses, and is able to better inform practical assessments of AMM pools.
\end{abstract}

\section{Introduction}

\newcommand{\nume}{num\'eraire\xspace}

\subsection{Exchanges and Liquidity Providers}
\label{subsec:intro_amms}

Decentralized exchanges (DEX's) are now an integral part of the broader ecosystem of blockchains, as evidenced in part by their ever growing volume of transactions.
The blooming popularity of DEX's in decentralized environments can be attributed to two prime factors:
first, the scarcity of available resources, and in particular storage and computation, so much that the dominant paradigm of exchange in traditional financial markets ---the central limit order book (CLOB)--- is essentially unimplementable on-chain,
and second, the need to provide liquidity and support price discovery in thin markets (markets with few buyers or sellers), for example, those of ``long-tail'' crypto assets.

The most common paradigm of decentralized exchange is that of automated market makers (AMMs) and, in particular, constant function market makers (CFMMs), with Uniswap \parencite{adams2021uniswap} being the most well-known and widely used example of a CFMM.
In AMMs, trades are enabled by the assets provided by liquidity providers (LPs) who act as market makers of the assets in the pool they participate in.
In particular, an LP pledges to the AMM some amounts of the assets being traded, and the AMM in turn provides rights to the fee income that comes from trading fees applied on traders.
At all times, traders may transact with the AMM at a marginal price that is defined as a function of the AMM's current reserves of those assets.
In this way, an AMM acts as a constant provider of liquidity to traders of either side of the market, buyers or sellers.
However, as the trade size becomes non-marginal, the AMM varies its provided price to the trader to diverge from the marginal one, in an inefficiency (compared to an infinite-depth or highly-liquid CLOB) that is widely known as ``slippage.''

By default, liquidity providers equally contribute to the AMM pool's liquidity irrespective of the asset price.
However, with the advent of the idea of concentrated liquidity \parencite{adams2021uniswap}, LPs can choose to only provide liquidity at a sub-range of asset prices, or even, by combining several so-called ``LP positions,'' vary their level of liquidity across the entire price spectrum.
This ability has significantly increased the possibilities of liquidity provisioning strategies by LPs, and has been conducive to encouraging them to more actively manage their demand curves dynamically as market conditions evolve, in a process that is akin to how positions in a CLOB are continuously adjusted.

\subsection{Adverse Selection Costs and Profitability in AMMs}
\label{subsec:intro_adverse}

In the process of providing liquidity to an AMM, LPs are unavoidably exposed to market risk due to the risky asset holdings that the AMM keeps on their behalf.
Additionally, as we mentioned in \Cref{subsec:intro_amms}, in order to facilitate trade, AMMs always accept (marginal) trades in either direction (buys or sells) at some quoted marginal price.
However, since AMMs are not aware of current asset prices, this necessarily implies an informational gap between informed traders (also called ``arbitrageurs'') who possess superior market information and the AMM.
What arises from this informational asymmetry between traders and liquidity providers is a phenomenon called ``adverse selection,'' that potentially undermines LP profitability stemming from the trading fee income stream they receive.
More specifically, arbitrageurs attempt to trade against the pool at every opportune moment in a zero-sum fashion, optimizing their trade such that they are always guaranteed a profit \parencite{jason_lvr_NEW}.
\textcite{jason_lvr} quantified this adverse selection cost as \LVR (loss-versus-rebalancing, pronounced ``lever''), which is now a popular metric to evaluate the toxicity of the incoming flow to an AMM.
\LVR is the only component that remains as a cost to LPs, if they continuously and fully hedge their market risk exposure to the risky assets their LP positions hold \parencite{jason_lvr}.

Besides informed traders (arbitrageurs), AMM pools also get trades from liquidity seeking agents who trade for at least partially idiosyncratic reasons (called ``noise traders'').
AMMs of course derive trading fee income from such traders, whose trades are uncorrelated with \emph{short-term} market price moves.
Thus, when evaluating LP performance \emph{ex post}, i.e., assessing past performance using realized data, one should consider the derived profit income stream as $\fee - \LVR$.

In general, LPs use various measures of adverse selection to inform their participation decisions.
Common metrics used by practitioners include markout and realized spread; see \textcite{huang1997components} for examples of such metrics.
Since the rise in popularity of AMMs, researchers have attempted to adapt conventional metrics to AMMs, and proposed new ones.
For instance, \textcite{thiccythot} uses markouts to evaluate the LP profitability in a concentrated liquidity pool of Uniswap protocol v3.

As a summary of the discussion above, it follows that flow toxicity stems from arbitrageurs (informed traders) transacting in a zero-sum way with the pool, whereas positive income stream comes from noise traders.
In traditional efficient markets, it is theoretically impossible to distinguish noise traders from informed traders ex ante under most commonly used and established economic models of behavior \parencite{kyle1989}; therefore, it may not be reasonable to attempt to resolve this issue in liquidity provisioning markets.

\subsection{Liquidity Providers' Competitiveness}
\label{subsec:intro_lp_competitiveness}

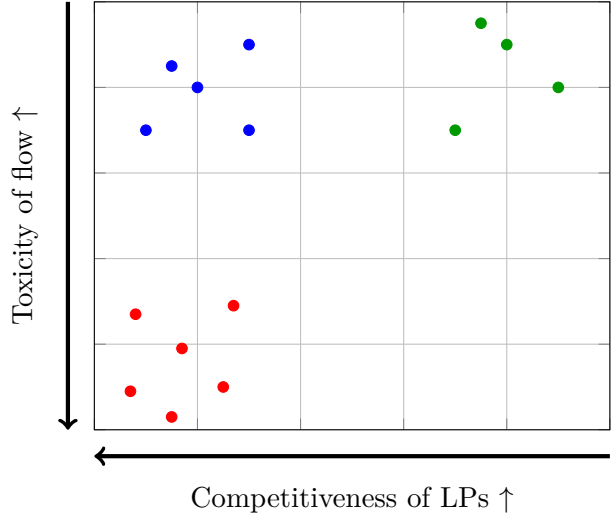
\begin{figure}[t]
    \centering
    \begin{tikzpicture}
    \begin{axis}[
    xmin=0,
    xmax=1,
    ymin=0,
    ymax=1,
    xlabel={},
    ylabel={},
    xticklabels={},
    yticklabels={},
    after end axis/.code={
    \draw [<-, ultra thick, below=10pt] (rel axis cs:0,0) -- (rel axis cs:1,0) node[midway, below=7pt] {Competitiveness of LPs $\uparrow$};
    \draw [<-, ultra thick, left=10pt] (rel axis cs:0,0) -- (rel axis cs:0,1) node[midway, rotate=90, above=7pt] {Toxicity of flow $\uparrow$};
    },
    grid=major,
    ]
    \addplot[ only marks, mark size=2, blue ] coordinates { (0.2,0.8) (0.3,0.9) (0.1,0.7) (0.3,0.7) (0.15,0.85) }; %
    \addplot[ only marks, mark size=2, black!40!green ] coordinates { (0.8,0.9) (0.7,0.7) (0.9,0.8) (0.75,0.95) }; %
    \addplot[ only marks, mark size=2, red ] coordinates { (0.08, 0.27) (0.25, 0.1) (0.15, 0.03) (0.07, 0.09) (0.17, 0.19) (0.27, 0.29) }; %
    \end{axis}
    \end{tikzpicture}
    \caption{Indication of what the assessment of aggregate LP returns of various pools would look like (synthetic data). The top right batch of pools has low flow toxicity and low competitiveness of LPs. The bottom left corner has pools with high flow toxicity and high competitiveness of LPs (very dynamic strategies, active LPing, JIT liquidity).}
    \label{fig:quadrant_intro}
\end{figure}

How much of that viable fee income stream is an individual LP able to capture, though?
In \Cref{subsec:intro_adverse}, we saw that an LP's performance is affected by its counterparty ---arbitrageur or noise trader.
Is this the only factor affecting the profitability of an LP?
Our answer with this paper is no, and this work aims to address this gap, and impact empirical studies of AMMs, as well as the way we fundamentally think about LP returns.

In particular, even though the delta-hedged fee income stream earned by a pool where a liquidity provider participates ---in the case that market risk is continuously and entirely hedged--- that arises as fees minus \LVR represents well and describes adequately the \emph{aggregate behavior} of the profitability of a liquidity pool, it is an incomplete descriptor of the returns of individual LPs, and does not translate well into an assessment of whether a new LP might be better off in one pool versus another.
This is precisely because there is an \emph{additional} factor to the toxicity of the fee income flow received by the pool that matters to individual LPs, one that is not captured by \LVR; this is the \textbf{competitiveness of liquidity providers}.
To make how the latter factor affects returns clear, let's begin by considering the following simple example as an indication of the insufficiency of flow toxicity to describe the \emph{micro-founded behavior} of a liquidity pool:
suppose that we have two AMM pools, A and B, that operate on the same assets, with uniformly distributed LP positions, and that at a time instant $t$ have exactly the same counterparty flow, i.e., fees minus \LVR is instantaneously the same for both pools.
LPs on pool A stay static during the next trade, and so do almost all LPs on pool B, except one particular LP who rushes to provide liquidity at the tightest possible range to the arriving trade (effectively simulating a request-for-quote (RFQ) system, otherwise known as providing just-in-time (JIT) liquidity).
On pool A, all LPs obtain uniformly the same share of fee income, but on pool B, it (almost) all goes to the LP that dynamically adjusted their position fastest and tightest compared to everyone else.
In aggregate, the income of the pool is \emph{exactly the same}, equal to the quantity ``fees minus \LVR'' that is the same in both pools.
However, at a \emph{micro-founded} level, on pool A, this income was uniformly distributed, whereas on pool B, it (almost) all went to one particular LP, and the other LPs obtained (almost) nothing.

It is, therefore, the case that an individual LP's returns critically depend on this notion of competitiveness among all of the liquidity providers in the pool, something that is not captured by the incoming flow to the pool, whatever the type of that flow is.
Currently, it is worth noting that very competitive strategies are not observed to be in place in the market \parencite{JIT_blogpost}, largely due to the early phase that decentralized finance is still in, but we predict that the landscape might change at any time with the growing maturity of the field.
Therefore, it would be beneficial for end users, as well as researchers, to have a metric that can account for the competitiveness of liquidity providers' positions on-chain, as well as the behaviors observed and the returns that were realized across a wide variety of pools.

To \emph{supplement} \LVR in assessing the \emph{dynamic behavior} of the LPs in a pool,
we introduce our LP competitiveness metric, FLAIR, short for fee liquidity-adjusted instantaneous returns.
FLAIR has multiple instantiations, depending on the exact target metric of interest (for a specific existing LP position, for the aggregate behavior of LPs in a pool, or for new LP positions), but is largely focused on capturing individual LP competitiveness characteristics.
A useful interpretation of such a metric in a manner that would be consistent with the above example and intuition can be based on what we call the \textbf{quadrant interpretation} of LP returns.
More specifically, referring to \Cref{fig:quadrant_intro}, there are two separate aspects / characteristics of LP returns that we generally wish to capture: 1. adverse selection (i.e., informational losses to arbitrageurs) which is captured by \LVR, and 2. active liquidity provisioning (i.e., dynamic LPing strategies, like just-in-time liquidity provision, that capture a significant proportion of the fee income stream) which we aim to capture in our metric.
In the examples given before, using just \LVR, it is not in general possible\footnote{Even though it might be in some cases, counterexamples can readily be constructed following the aforementioned presented reasoning.} to distinguish between high fee-return-on-capital and low fee-return-on-capital pools.
In particular, in \Cref{fig:quadrant_intro}, (unsophisticated) LPs should generally prefer to participate in green pools rather than in blue pools, even though both of them exhibit low adverse selection costs (\LVR), because the necessary strategy to employ in such pools is easier to deploy and follow.

In this work, we intend to be examining the competitiveness of LPs in an \emph{ex-post} fashion, i.e., assessing the realized past performance within some time frame.
This means that some specific price trajectory was realized, combined with specific liquidity provision strategies by each individual LP that participated in any given pool at any given time instant.
In turn, the fact that there was a specific price trajectory implies that, according to \Cref{subsec:intro_adverse}, there is a good candidate of a measure for adverse selection costs (otherwise referred to as toxicity of the flow) that is the \emph{realized} instantaneous rate of \LVR.
Continuing from that point, we formally define in \Cref{subsec:metric_existing_lps} the first instantiation of FLAIR, our LP competitiveness metric, that matches the intuition given above and focuses on individual LP returns.
More formally, the FLAIR for an individual LP in a pool amounts to the fee return on capital that this individual LP position was able to capture.
Inside a liquidity pool, this means the proportion of fees that correspond to a particular LP, normalized by the total provided funding capital of this position at each time.
In other words, if the market value of the portfolio holdings of LP position $i$ at time instant $t$ is denoted by $V_i(t)$, then the FLAIR for this position will be defined in \Cref{eq:cm_i} as (see \Cref{subsec:metric_existing_lps} for the full details and notation)
\begin{equation*}
\cm_i(t_0, T)
\triangleq
\int_{t_0}^T
\frac{1}
{V_i(t)}
\times
\fee_t
\times
\frac{L_i(\tilde{p}_t;t)}{L(\tilde{p}_t;t)}
\, dt
\,.
\end{equation*}

Finally, we would also be interested in exhibiting a version of FLAIR in the case that a \textbf{new LP} is considering their ex-post participation in a pool, and wants to assess the potential fee return on capital they would have, had they followed the best strategy they could have out of the ones they have available (i.e., backtesting liquidity provisioning strategies).
To do so, we employ a definition of a set $\calL$ of available LP strategies that is dependent on what the LP considers feasible for them to deploy, and we formulate an optimization problem based on this general definition that resembles those familiar from modern portfolio theory.
For example, it would make sense for unsophisticated LPs to think about the passive (i.e., not changing ``frequently'' in time) liquidity provisioning strategies they could decide (i.e., optimize) once about, and then revise only coarsely often.
More generally, the strategy pointed at by maximizing LP competitiveness is the one that attempts to compete \emph{all the time} with the other LPs in a pool (leading to the capturing of the most market share of fees), irrespective of the received order flow.
Notice that, because order flow might be extremely adversarial at times, this might not lead to the strategy maximizing competitiveness being the same as the strategy maximizing profitability.
This is why we show a minor variation of our optimization problem---which intuitively combines FLAIR with the LP's preferred order flow toxicity metric, such as instantaneous LVR---that allows the prospective LP to \emph{optimize precisely the profitability} of the employed strategy.
Therefore, a strong connection to profitability is obtained in all cases.
We leave the details for \Cref{subsec:metric_new_lps}.

On the whole, FLAIR is an important notion to have, one that is orthogonal to LVR, to better account for LP returns, because it allows the \emph{dissection} of LP returns into a component due to the order flow toxicity and a separate one due to the competitiveness of LPs.
In sharp contrast, markout and other sheer aggregate profitability metrics conflate these two notions into one.
The specific attribution of LP returns to either one or the other has implications on selecting among the venues of offering liquidity for the same asset pairs, versus the specific asset class to provide liquidity on.

\subsection{Related Work}
\label{subsec:litrev}

Despite the fact that studies of liquidity provisioning on constant product market makers as well as concentrated liquidity Uniswap protocol v3-like AMMs have been abundant (see \textcite{cohen2023inefficiency,bergault2022automated,fan2022differential,huynh2022providing,neuder2021strategic,yin2021liquidity} for just a few of them), to the best of our knowledge, there has been no prior principled theoretical study of liquidity providers' competitiveness among each other \emph{irrespective} of specific characteristics of particular CFMM implementations or AMM curves.
This work is the first to offer such a novel perspective into the competitiveness of LPs and pools.

The study of automated market makers as decentralized exchange mechanisms has multiple reference points, with convex analysis tools having been used in analyses by \textcite{angeris2020improved,angeris2021replicatingmarketmakers,angeris2021replicatingmonotonicpayoffs} to, among others, define the optimal portfolio value after arbitrageur rebalancing and replicate payoff functions.
We use portfolio values in this work, as well as a generalization (that we first define) of an alternative representation of a CFMM curve (established by \textcite{angeris2021replicatingmarketmakers,angeris2021replicatingmonotonicpayoffs}) in terms of asset holdings of the pool as a function of the pool's implied (spot/marginal) price.
Based on the latter representation, \textcite{jason_revenue_optimal_LP,jason_exchange_complexity} introduce a novel general framework for market making (what they term as an ``exchange mechanism'') that is able to elucidate and disambiguate the \emph{individual} incentives of LPs, including the differences between them.
\textcite{jason_revenue_optimal_LP} then supplement this framework with Bayesian-like belief inference, where LPs maintain an asset price estimate, which is updated by incorporating traders' price estimates and that LP's prior beliefs about asset prices.

On the LP returns front, \textcite{jason_lvr} were the first to decompose the returns of liquidity providers into an instantaneous market risk component and a non-negative, non-decreasing, and predictable component called ``loss-versus-rebalancing'' (\LVR, pronounced ``lever'') using a continuous-time Black-Scholes setting in the absence of trading fees for arbitrageurs. They showed that the market risk is fully hedgable, and what remains after removing it is exactly the adverse selection cost due to the arbitrageurs' informational advantage to the pool.
Our LP competitiveness metric \emph{supplements} \LVR in assessing the dynamic behavior of the LPs in a pool.

\subsubsection{Traditional Finance}
\label{subsubsec:rel_work_tradfi}

Traditional finance literature provides rich insights into the operations of liquidity providers (LPs), thereby informing our analysis of LP competitiveness in automated market makers. Seminal work by \textcite{ho1981optimal, ho1983dynamics} laid the foundation for understanding dealer behavior and market making, introducing models of optimal dealer pricing under transaction and return uncertainty. Subsequent research, including by \textcite{madhavan1993analysis, hansch1998inventories}, further explored LP behavior, linking inventory management and market conditions to profitability. Moreover, studies like \textcite{bessembinder2004does, foucault2003market} underscore the influence of market structure on LP strategies, with implications for our understanding of AMMs.

\section{Metric Theory}
\label{sec:metric}

\subsection{Preliminaries}
\label{subsec:metric_prelim}

Assume a two-asset AMM pool, where, without loss of generality, the first asset $A$ is a risky asset, and the second asset $B$ is a \nume asset (the unit of account), in the units of which the various portfolio values, fees, profit and loss will be denominated.

Denote by $p_t$ the external market price of the risky asset. Whenever our measures are ex-post, this means that there exists a specific realized price trajectory $\{p_t\}_{0\le t\le T}$, and that we are measuring the past performance against that specific sample path.

An AMM in the form of a constant function market maker (CFMM) is represented by a bonding curve $f$ such that the only allowable trades are those that maintain a constant level set of the curve, i.e., $f(x,y)=L$.
Through this representation, at any state of the pool $(x,y)$, under suitable technical assumptions, we can define a marginal price, the so-called implied pool (marginal/spot) price that is given by the negative of the ratio of the partial derivatives of $f$ evaluated at the point $(x,y)$.

We denote the implied pool (marginal/spot) price by $\tilde{p}_t$. This depends solely on the pool state (the existing reserves and liquidity in the pool).
In the case that arbitrageurs trade continuously with no trading fees in the AMM, the optimal solution of the portfolio value optimization problem first formulated by \textcite{angeris2021convopt} shows that $\tilde{p}_t = p_t$ at each time instant.
On the other hand, the implied pool price may at times diverge from the external market price $p_t$ if arbitrageurs trade with trading fees on the AMM \parencite{jason_lvr_NEW}\footnote{For developing the theory of our competitiveness metric, there is no need to assume that arbitrageurs rebalance the pool until the external market price, i.e., as if they would pay no fees for their trades. Our theory works in the completely general case where there is mispricing between the implied pool price $\tilde{p}_t$ and the external market price $p_t$.}.

\paragraph{Reparameterization of the bonding curve in the price space.}
One particularly useful reparameterization of a CFMM curve that, in this work, we will find helpful to think in terms of is the ``price space'' reparameterization.
More specifically, given a bonding curve $f$ that defines a CFMM, the following are equivalent: specifying the reserve quantities $(x,y)$ is equivalent to specifying the pair $(p,L)$ of the implied (marginal/spot) pool price\footnote{Observe that this reparameterization readily allows the incorporation of trading fees on arbitrage trades, since the implied pool price may at times deviate from the external market price.} combined with the current level set of the bonding curve.
Both directions of the equivalence are easy to see, hinge on the implicit function theorem \parencite{krantz_implicit_2003} under suitable smoothness assumptions for the involved bonding curve $f$, and have been exhibited before in various forms in prior work on replicating payoffs in asset portfolios (see, for example, the works of \textcite{angeris2020improved,angeris2021replicatingmonotonicpayoffs}).
In particular, in this work, we will denote $x^\star(p, L)$ and $y^\star(p, L)$ for this reparameterization, i.e., the function that take a point in the price space $(p,L)$ and transform it to a point in the reserves space $(x,y)$.

\paragraph{Liquidity distribution.}
Traditionally, in constant function market makers, trades are allowed to occur if and only if the resulting asset reserves are located on a (constant) level set of the bonding curve, i.e., the reserves $\left( x, y \right)$
must satisfy $f\left( x, y \right) = L$ for a specified constant $L$ (the level).
With the advent of concentrated liquidity, the idea originating from Uniswap protocol v3 has been that the price space may be decomposed into discrete intervals such that the liquidity constant $L$ can be allowed to vary from interval to interval.
For an extremely simple example, a price interval $[p_0, p_2]$ may be decomposed into two price intervals $[p_0, p_1]$ and $[p_1, p_2]$ such that the reserves of the pool satisfy within the first interval the level set $f\left( x, y \right) = L_1$ when $p \in (p_0, p_1)$ and $f\left( x, y \right) = L_2$ when $p \in (p_1, p_2)$, for potentially different constants $L_1 \ne L_2$.

To generalize this idea, we use the reparameterization described in the previous paragraph as follows: define arbitrary (discrete, potentially infinitely many, and arbitrarily fine) price intervals of the form $[p_k, p_{k+1}]$ for indices $k\in\calI$ of some generic indexing set $\calI$ (for instance, $\N$).
In particular, we say that there exists a piecewise-constant \textbf{liquidity distribution} $L(p)$
that has the property that $L(p) = L_k$ for all $p\in (p_k, p_{k+1})$ for any $k\in\calI$
such that
whenever the implied (marginal/spot) pool price $p\in (p_k, p_{k+1})$, the valid reserves $(x,y)$ contained in the pool satisfy the relationships $x = \xs(p, L(p))$ and $y = \ys(p, L(p))$ for all $p\in (p_k, p_{k+1})$ for any $k\in\calI$.

For example, in the case of Uniswap protocol v3, the respective functions would be\footnote{Notice that the reserves depend on the entire liquidity distribution function $L(p)$ and not only on the current liquidity $L_k$, because we are not only referring to the current reserves in the interval $(p_k, p_{k+1})$ but are including the \textit{total} reserves present in the pool, even due to the other (passive/non-active) intervals.}
\begin{align*}
\xs(p, L(p))
&=
L_k \cdot \left( \frac{1}{\sqrt{p}} - \frac{1}{\sqrt{p_{k+1}}} \right)
+
\sum_{m > k}
L_m \cdot \left( \frac{1}{\sqrt{p_m}} - \frac{1}{\sqrt{p_{m+1}}} \right)
\text{ and }
\\
\ys(p, L(p))
&=
L_k \cdot \left( \sqrt{p} - \sqrt{p_k} \right)
+
\sum_{m < k}
L_m \cdot \left( \sqrt{p_{m+1}} - \sqrt{p_m} \right)
\ \forall p\in (p_k, p_{k+1}),\, k\in\calI
\,.
\end{align*}

\subsection{FLAIR for Existing LP Positions}
\label{subsec:metric_existing_lps}

As mentioned in the introduction, the ex-post competitiveness metric for existing LP positions should assess the backwards-looking profitability of the position, measured by the fee earned in comparison to the rest of the pool, normalized by the instantaneous amount of capital deployed on the AMM, as this position varies its liquidity distribution over time.
In other words, the metric measures the \emph{instantaneous rate of fee return on the capital deployed in the AMM} through the $i$-th position's time-varying liquidity distribution, integrated over a time interval $[t_0, T]$ that we would like to assess the metric on.
The first instantiation of FLAIR (fee liquidity-adjusted instantaneous returns) is thus going to be, as we will see below, $\cm_i(t_0, T)$.

In particular, let's assume that there exist positions/LPs\footnote{In the theory, we use the two notions interchangeably, but as far as the empirical results are concerned, such a nuance would make a meaningful difference.} in a pool such that the $i$-th position/LP has supplied at the time instant $t$, when the implied pool price is $\tilde{p}_t$, a liquidity distribution $L_i(p;t)$ in the pool\footnote{Notice that the index $i$ here does not refer to the same meaning as in \Cref{subsec:metric_prelim}, where the index referred to the $k$-th price interval; here, it refers to the entire liquidity distribution function of the $i$-th position/LP. Also notice the time-varying aspect of the liquidity distribution; we allow arbitrary changes to it, which can capture arbitrary discrete block times, for example.}.
Note that, as mentioned in \Cref{subsec:metric_prelim}, a liquidity distribution is an \emph{entire function} of price (and \emph{not} a single value that arises from plugging in the current implied pool price), because the deployed liquidity varies with price.
In total, the aggregate liquidity distribution in the pool is $L(p;t) = \sum\limits_{i} L_i(p;t)$.

Denote by $\fee_t$ the instantaneous fee rate earned by the \textit{entire pool} due to all trades (i.e., both noise and arbitrage trades) at time instant $t$; in other words, $\int_0^T \fee_t\, dt$ are the total fees (denominated in \nume units) earned by the entire AMM pool until time $T$.
Finally, we note that, when the external market price of the risky asset is $p_t$, and the implied pool price is $\tilde{p}_t$ (where $\tilde{p}_t=p_t$, if the arbitrageurs are assumed to always trade until the external market price, irrespective of the trading fee, otherwise differing according to some bounded mispricing process), then the market value of the portfolio holdings of position $i$ at time instant $t$ will be $V_i(t) \triangleq p_t \cdot \xs\left( \tilde{p}_t, L_i(p; t) \right) + \ys\left( \tilde{p}_t, L_i(p; t) \right)$.

Since we are measuring the instantaneous fee return on capital deployed, the instantaneous metric needs to be normalized with the portfolio value of the position at time $t$.
Additionally, we remark that the fee earned by the position is proportional to its active/in-range deployed liquidity, compared to the entire pool's active/in-range deployed liquidity.

Following the aforementioned notation, we define FLAIR for the existing LP position $i$ as\footnote{In what follows, we can set $t_0=0$ to obtain the entire metric until time $T$, but for generality, we include the expressions for an arbitrary $t_0<T$ hereby.}
\begin{equation}
\label{eq:cm_i}
\cm_i(t_0, T)
\triangleq
\int_{t_0}^T
\frac{1}
{V_i(t)}
\times
\fee_t
\times
\frac{L_i(\tilde{p}_t;t)}{L(\tilde{p}_t;t)}
\, dt
=
\int_{t_0}^T
\frac{\fee_t}
{p_t \cdot \xs\left( \tilde{p}_t, L_i(p; t) \right) + \ys\left( \tilde{p}_t, L_i(p; t) \right)}
\times
\frac{L_i(\tilde{p}_t;t)}{L(\tilde{p}_t;t)}
\, dt
\,.
\end{equation}

\subsubsection{Aggregate Pool Competitiveness Metric and Competitiveness Quadrant}

One reasonable question is how we would aggregate the FLAIR for all LP positions into one single aggregate metric for the entire pool, which would indicate, over time, what the fee return on capital deployed in that pool is.
In particular, the aggregate FLAIR for the pool that would measure such a fee return should be computed by a similar formula to the above one, i.e., the ex-post aggregate pool competitiveness metric should be
\begin{equation}
\label{eq:cm_agg}
\cm_\text{agg}(t_0, T)
\triangleq
\int_{t_0}^T
\frac{1}
{V(t)}
\times
\fee_t
\, dt
=
\int_{t_0}^T
\frac{\fee_t}
{p_t \cdot \xs\left( \tilde{p}_t, \sum\limits_i L_i(p; t) \right) + \ys\left( \tilde{p}_t, \sum\limits_i L_i(p; t) \right)}
\, dt
\,.
\end{equation}

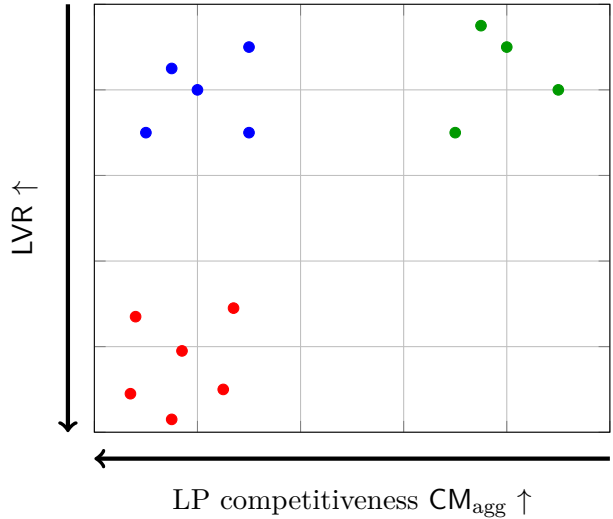
\begin{figure}[H]
    \centering
    \begin{tikzpicture}
    \begin{axis}[
    xmin=0,
    xmax=1,
    ymin=0,
    ymax=1,
    xlabel={},
    ylabel={},
    xticklabels={},
    yticklabels={},
    after end axis/.code={
    \draw [<-, ultra thick, below=10pt] (rel axis cs:0,0) -- (rel axis cs:1,0) node[midway, below=7pt] {LP competitiveness $\cm_\text{agg}$ $\uparrow$};
    \draw [<-, ultra thick, left=10pt] (rel axis cs:0,0) -- (rel axis cs:0,1) node[midway, rotate=90, above=7pt] {\LVR $\uparrow$};
    },
    grid=major,
    ]
    \addplot[ only marks, mark size=2, blue ] coordinates { (0.2,0.8) (0.3,0.9) (0.1,0.7) (0.3,0.7) (0.15,0.85) }; %
    \addplot[ only marks, mark size=2, black!40!green ] coordinates { (0.8,0.9) (0.7,0.7) (0.9,0.8) (0.75,0.95) }; %
    \addplot[ only marks, mark size=2, red ] coordinates { (0.08, 0.27) (0.25, 0.1) (0.15, 0.03) (0.07, 0.09) (0.17, 0.19) (0.27, 0.29) }; %
    \end{axis}
    \end{tikzpicture}
    \caption{Quadrant interpretation of LP returns (synthetic data).}
    \label{fig:quadrant}
\end{figure}

As described in \Cref{subsec:intro_lp_competitiveness}, we would like to decompose the LP returns into a component that has to do with the toxicity of the flow that the pool receives, and another one that is relatively ``orthogonal'' to the first, representing the competitiveness of the LP positions in the pool.
We now have naturally motivated metrics for both.
As such, we have what we call the \textbf{quadrant interpretation} of LP returns, which is concretely shown in \Cref{fig:quadrant}\footnote{Note that the figures amount to synthetic data, and we did not perform empirical analysis of actual pools to generate the data. As this is ongoing work, our current direction is performing empirical analysis based on our metrics.}.
In this figure, which is substantially the same as \Cref{fig:quadrant_intro}, except now with completely specified axes, we are therefore going to use \LVR for the toxicity of the flow, and for the competitiveness of the pool we are proposing to use $\cm_\text{agg}$ as defined in \Cref{eq:cm_agg}.

\subsection{FLAIR for New LP Positions}
\label{subsec:metric_new_lps}

The purpose with establishing a FLAIR for new LP positions is for this competitiveness metric to be an indicator of the potential profitability of a new LP position that is capital-constrained and is considering the deployment strategy of its capital for liquidity provision.
The metric will still be ex-post, i.e., it will use past data to account for how a new LP position would have behaved in the entire time frame until a time $T$\footnote{As with \Cref{subsec:metric_existing_lps}, it might make sense to consider the metric from some time moment $t_0 \ne 0$ up to $T$. In an aggregate pool metric, we posit that such a consideration is even more important than before, because of how the competitiveness of a pool might vary dynamically, hence we might not desire the metric to be affected by potentially dated data. For instance, one might be interested in examining the times in the last month/day/minute prior to a current time $T$.}.
More specifically, in this case, in order to propose a metric, we first have to think about the potential liquidity provision strategies that may be employed by this new position.

To this end, consider the \textbf{set of all pre-defined liquidity provision strategies} $\calL$ to be the set of functions $L_\text{new}(p; t)$ that are candidates that the LP is considering for the new position, and might act according to any one of them.
Such strategies could potentially depend on a variety of factors, ranging from the current external market price, the current implied pool price\footnote{For instance, trying to preempt arbitrageurs might be one such highly-competitive strategy.}, and the current fee rate to what other liquidity provisioning strategies do in the pool.
On the other end, these strategies could potentially be as simple as only encompassing passive liquidity provision, i.e., making an initial decision at the start of capital deployment $t_0$ and then following this steadily with no changes for the entire rest of the LPing time period. Such a restrictive set $\calL$ would correspond to any liquidity provision strategy $L_\text{new}(p; t) \in \calL$ being allowed to only be of the form $L_\text{new}(p; t) = L_0(p)$, i.e., independent of time.

Now that we have in place this generic set of allowable liquidity provisioning strategies $\calL$ as described, the competitiveness of a new LP position would then be the maximum competitiveness that this position could achieve, using a fixed starting portfolio (capital-constrained at some amount $c$) valued at the initial time instant $t=t_0$, using any of its feasible potential liquidity provision strategies defined by the set $\calL$.
Such a definition is natural, because it is expected that the liquidity provisioning strategy that would be utilized by a new position would be, among the allowable ones, the one that grants the maximum competitiveness (best-response) with respect to what the pool is doing in aggregate.

Using the above knowledge and \Cref{eq:cm_i}, we define the FLAIR of a new LP position as\footnote{Note that in this equation, the aggregate liquidity in the pool $L(p; t)$ needs to include not only the current strategies in the pool, but also the new one, $L_\text{new}(p; t)$.}
\begin{align}
\label{eq:cm_new}
\cm_\calL(c; t_0, T)
\triangleq
\sup_{L_\text{new} \in \calL}
&\cm_{L_\text{new}}(t_0, T)
\\ \text{s.t.} \quad
&V_{L_\text{new}}(t_0) = c
\,.
\nonumber
\end{align}

Fully writing down the above optimization problem using the metric \Cref{eq:cm_i} of \Cref{subsec:metric_existing_lps} would yield our following proposed complete competitiveness metric for new LP positions:
\begin{align*}
\cm_\calL(c; t_0, T)
\triangleq
\sup_{L_\text{new} \in \calL}
&\int_{t_0}^T
\frac{\fee_t}
{p_t \cdot \xs\left( \tilde{p}_t, L_\text{new}(p; t) \right) + \ys\left( \tilde{p}_t, L_\text{new}(p; t) \right)}
\times
\frac{L_\text{new}(\tilde{p}_t; t)}{L(\tilde{p}_t; t)}
\, dt
\\ \text{s.t.} \quad
&p_{t_0} \cdot \xs\left( \tilde{p}_{t_0}, L_\text{new}(p; t_0) \right) + \ys\left( \tilde{p}_{t_0}, L_\text{new}(p; t_0) \right) = c
\,.
\end{align*}

An important observation is that the above metric optimizes the chosen strategy based on trying to capture the \emph{market share} of the fees, adjusted for the available capital.
However, as is well known, this may not always lead to the optimal profitability of the corresponding position; the reason is that many times, whenever the fees are (extremely) high, asset price volatility is high, which induces a high degree of adverse selection on the LP.
As mentioned before, this has to do with the order flow toxicity, which we would like to be a separate, orthogonal measure; nonetheless, when optimizing for a portfolio choice or liquidity provisioning strategy, the optimization of the two measures may be in contrast.
In this case, the above analysis means that if, instead of a pure metric of competitiveness (that strives to \emph{always} compete with the present LPs on-chain, capturing the most market share of fees), a \emph{pure} profitability metric is to be desired, then the metric in point should, instead of $\fee_t$ in the above formula, have $\fee_t$ minus the preferred order flow toxicity metric of choice that the LP prefers (whether that is instantaneous LVR $\ell_t$, or a generalized notion of markout).
The optimization problems then changes correspondingly, and as an example, if we used instantaneous LVR, it would have been
\begin{align*}
\sup_{L_\text{new} \in \calL}
&\int_{t_0}^T
\frac{\fee_t - \ell_t}
{p_t \cdot \xs\left( \tilde{p}_t, L_\text{new}(p; t) \right) + \ys\left( \tilde{p}_t, L_\text{new}(p; t) \right)}
\times
\frac{L_\text{new}(\tilde{p}_t; t)}{L(\tilde{p}_t; t)}
\, dt
\\ \text{s.t.} \quad
&p_{t_0} \cdot \xs\left( \tilde{p}_{t_0}, L_\text{new}(p; t_0) \right) + \ys\left( \tilde{p}_{t_0}, L_\text{new}(p; t_0) \right) = c
\,.
\end{align*}
Finally, notice that this metric now gives us an \emph{exact} correspondence between the solution of this optimization problem and the profitability of the employed strategy, which is precisely this optimal computed point.
The metric, in this case, would simply capture the entire returns of the LPing portfolio achieved through the best of the strategies chosen.

\section{Examples of the Metric}
\label{sec:metric_ex}

In this section, we present a few cases that show how FLAIR varies in different scenarios that are intuitively expected to be more or less competitive.
The purpose of this section is to showcase that our metric aligns with what would normally be expected to be characterized as ``competitiveness'' among LPs in a pool indicating that our formalization of this notion through our metric in \Cref{sec:metric} has matching natural interpretations.
For simplicity, we present here only the ex-post aggregate pool competitiveness metric, since this is enough to exhibit our point.

The examples we present all follow the below common premises, and differ on the key aspects that will be discussed in their respective subsections:
\begin{itemize}
    \item We consider that there are only 2 ``equivalent'' positions in the pool, one of which will be our ``test'' position (the one on which we are going to evaluate the metric) and the other one which will represent the aggregate of all other positions in the pool. %
    We assume that each of these positions carries 50\% of the capital.
    \item There is a constant fee rate\footnote{Notice that this does not mean constant fees for each LP, because of the potentially dynamically employed active liquidity provisioning strategies.} $\fee_t = f = 1$.
    \item For the realized price trajectory, we are going to examine two cases: in the first case, the price remains constant throughout the examined time period, and in the second case, it will be linearly increasing from $t_0$ to $T$, such that $p_t=\tilde{p}_t=\pmin + (\pmax-\pmin) \cdot \frac{t-t_0}{T-t_0}, \ \forall t\in[t_0,T]$.
\end{itemize}

\subsection{CFMMs}

In the case of a traditional constant function market maker (CFMM), where the entire curve does not exhibit concentrated liquidity, all LP positions have to be full-range, providing liquidity over the entire price range $(0, \infty)$.

This restricts the set of possible LPing strategies on a CFMM to one strategy, which forces both positions to utilize the same one. Two fully competitive LPs will place their liquidity into the pool at time $t_0$ and make no adjustments, because adjusting cannot increase their competitiveness in the pool. In the example case of the price trajectory being constant, both LPs utilize $V(t)=c$ for some $c$, therefore this will result in the aggregate competitiveness of the pool equaling
\[
\cm_\text{agg}(t_0, T)
=
\int_{t_0}^T
\frac{\fee_t}
{2c}
\, dt
\,
=
\frac{\int_{t_0}^T \fee_t \, dt}
{2c}
=
\frac{T - t_0}{2c}
\,.
\]

More generally, this simple starting case confirms the obvious observation that in simple CFMMs where no active LPing is allowed and all positions are full-range, the instantaneous competitiveness of all positions remains the same through time, conditioned on them not varying the relative amount of capital deployed in the pool (i.e., the fraction $\frac{L_1}{L_1+L_2}$).
In other words, competitiveness of LPs has no effect on simple CFMMs, which is the expected behavior. This is because there is (trivially) no active strategy that can be employed on such CFMMs, apart from selecting whether to deploy capital for liquidity provision or not and how much capital, something that is ---in aggregate--- being abstracted away due to the normalization of return-on-capital.

\subsection{Uniswap protocol v3 with fully competitive LP positions}

Next, we examine the case of a concentrated liquidity AMM, where (some or all) LP positions attempt to constantly track as narrowly as possible the price, i.e., intuitively pools where ``full competitiveness'' of (some or all) positions is expected to be observed.
For example, such a situation is commonly observed in Uniswap protocol v3 ETH-USDC 5 bp pools. The flow of this pool is dominated by competitive LPs, who place concentrated orders, realizing significant fees for comparatively little underlying capital.

\subsubsection{All fully competitive positions with constant price}

At the limit of competitiveness, two perfectly competitive LPs should place only the exact amount of capital needed to trade against the flow at each time instant.
Assuming that $\gamma$ is the trading fee (as a proportion of the total trading volume) of the pool and $n$ is the amount of LPs in the pool, and since we have two ($n=2$) LPs that provide 50\% of the liquidity of the pool, both LPs will individually provide at time $t$ capital of
\[
V_i(t)
=
\frac{2\fee_t}{n\gamma}
=
\frac{\fee_t}{\gamma}
\,.
\]

Notice that the reason that the numerator is multiplied by 2 is non-trivial, and related to trading mechanics and the CPMM in particular. This is because the LPs need to provide a portfolio of both tokens in the current tick to facilitate trading, and we assume that a proportional fee of $\gamma$ is taken from \emph{both} traded tokens. In practice, the multiplier might be slightly more or less than two times the value of the flow depending on the placement of the tick within the tick-range. However, for simplicity, we will assume that the tick is perfectly in the middle of the tick-range, allowing the perfectly competitive LPs to place equal portfolio values in the two tokens.

Such is then the amount of capital needed to perfectly trade against all flow that results in the fee income stream of $\fee_t$ with no excess capital deployed.
As a direct consequence, the aggregate FLAIR for the pool would be
\[
\cm_\text{agg}(t_0, T)
=
\int_{t_0}^T
\frac{\fee_t}
{2\frac{\fee_t}{\gamma} }
\, dt
=
\int_{t_0}^T
2\gamma
\, dt
=
\
2\gamma(T - t_0)
\,.
\]

\subsubsection{All fully competitive positions with varying price trajectory}

Because the two LPs will perfectly mirror each other, and both LPs instantaneously adjust their positions to the react to the external price changes,\footnote{This indeed presupposes that the LPs adjust their positions faster than adversarial traders (arbitrageurs) might arrive to take advantage of the price movement, up to the narrowest possible tick-range. As a reminder, these examples are not indicative of real-world behavior, but showcase the interpretation of various LP competition scenarios, and as such, do not focus on what exact empirical analyses would show.} their FLAIR as well as the aggregate FLAIR of the pool will not change.
This is because they will both perfectly and instantaneously place capital needed to swap against the flows of the pool.

\subsubsection{One low-competitiveness position with passive liquidity}

Finally, consider the case of a Uniswap protocol v3 pool where one position provides passive liquidity, and the other (aggregate, in the sense described above) LP is fully-competitive. In this case, the expectation is that, while the pool as a whole is very competitive, this particular passive position exhibits low competitiveness, and would thus be particularly harmed in this pool, even if it exhibited low adverse selection costs.

First, we must introduce the notion of the tick-spacing, $ts$ in the Uniswap protocol v3 pool. The tick-spacing is the minimum distance between concurrent ticks where liquidity can be placed. This in turn enforces the maximum concentration of an LP position, as the maximum concentration occurs at the minimum distance between the start and end ticks of a position. Because fully competitive LPs will always concentrate their liquidity over the smallest range, the tick-spacing is the main driver of the liquidity differences between a fully competitive and a passive LP.

Let both LPs utilize $L$ liquidity. The passive liquidity provider observes the price path and places their liquidity starting at above $\pmin$ and ending at below $\pmax$.

However, the placements would not be exactly these, because LPs cannot generally place liquidity directly on these prices. This is because of the minimum distance required between ticks by the tick-spacing. Because of this, they must calculate the minimum upper tick, $t_u(p)$ to place above price $p$ and maximum lower tick $t_l(p)$ below price $p$, which is dependent on the tick-spacing. This can be done by the following equations:
\begin{align*}
    t_u(p) &= 
    \lceil 
    \frac{\log_{1.0001} p}{ts} 
    \rceil 
    \cdot 
    ts
    \,, \text{ and}
    \\
    t_l(p) &=
    \lfloor
    \frac{\log_{1.0001} p}{ts} 
    \rfloor
    \cdot 
    ts
    \,.
\end{align*}

Each tick $t$ can then be translated to the price by the calculation $1.0001^{t}$.
The underlying tokens needed for the passive liquidity provider would then be
\[
    c_p = L(\sqrt{1.0001^{t_u(p_{max})}} - \sqrt{1.0001^{t_l(p_{min}}))}
\]

The fully-competitive LP is more complicated with the varying price trajectory. In particular, they observe the price of the pool at time $t=t_1$ and $t=t_2$, and find $p_1 = \pmin + (\pmax-\pmin) \cdot \frac{t_1-t_0}{T-t_0}$ and $p_2 = \pmin + (\pmax-\pmin) \cdot \frac{t_2-t_0}{T-t_0}$. 

With all of this, we can calculate the amount of tokens needed for the positioning of the competitive LP, $c_c$ by
\[
    c_c = L(\sqrt{1.0001^{t_u(p_2)}} - \sqrt{1.0001^{t_l(p_1)}})
\]

Finally, we compute the aggregate FLAIR of the pool as in the following calculation:
\[
\cm_\text{agg}(t_0, T)
=
\int_{t_0}^T
\frac{\fee_t}{
L(\sqrt{1.0001^{t_l(p_{t+1})}} - \sqrt{1.0001^{t_l(p_t)}}
+ \sqrt{1.0001^{t_l(p_{max})}} - \sqrt{1.0001^{t_l(p_{min})}})
}
\, dt
\,.
\]

Unfortunately, in general, such a calculation would be heavily dependent on the tick-spacing and rounding errors due to the price path of the pool.
This is why we focus instead on the following observations:
\begin{itemize}
    \item If the price bounces sufficiently back and forth between the maximum and the minimum possible price, then the most efficient liquidity provision is full-range.
    \item If the tick-spacing of the pool $ts$ is sufficiently large enough that $t_u(\pmin) \ge \pmax$, then the same calculation converges to the case of the two sophisticated/fully-competitive LPs examined previously. Another example of this is a constant price path as in both cases the LPs do not adjust their positions: in the latter case, this is due to the price never moving, and in the former case, this is due to the tick-spacing not being narrow enough to allow adjustments (i.e., enough concentration of capital). %
\end{itemize}

\section{Further discussion}

We conclude with some further discussion of observations around FLAIR and LP competitiveness.

First, FLAIR is well-defined for individual positions point-in-time, as well as over arbitrary periods, and for arbitrary collections of positions, such as a portfolio or an entire pool. This allows both existing LPs to measure historical performance, and prospective LPs to back-test and optimize for future capital deployments. Researchers can also categorize different strategies and LPs based on their FLAIR. Solving for the ``optimal frontier'' will be an interesting portfolio optimization question.

Second, under this metric, the comparative performances of many well-analyzed scenarios and strategies reflect reasonable economic intuition. Holding other things equal, in-range liquidity concentration increases instantaneous competitiveness, allocation to pools with higher fee returns increases competitiveness, and better timing of deployment to high fee periods increases competitiveness. On a forward-looking basis, a tighter constraint on rebalancing frequency, as well as a larger capital amount, will lead to a wider optimal range.

Finally, FLAIR could be applied, with some terminological modification, to traditional exchanges as well, allowing comparative studies between different market structures and influencing future market designs.

\section*{Acknowledgments}
We thank Ciamac C. Moallemi for engaging discussions on a variety of different versions of our metric and models, as well as on interpretations of our metric.

\printbibliography

\end{document}